\documentclass[fleqn,usenatbib]{mnras}

\usepackage{newtxtext,newtxmath}

\usepackage[T1]{fontenc}
\usepackage{ae,aecompl}

\usepackage{graphicx}	
\usepackage{amsmath} 	
\usepackage{mathtools}  

\defcitealias{kittiwisit.etal.2018}{K18}
\newcommand{\wcbw}{\ensuremath{\Delta \nu_{\mathrm{WC}}}}

\title[Foreground wedge-cut 21 cm one-point statistics]
{Measurements of one-point statistics in 21 cm intensity maps via foreground avoidance strategy}
\author[P. Kittiwisit et al.]{Piyanat Kittiwisit$^{1,2}$\thanks{Contact e-mail: \href{mailto:piyanat.kittiwisit@gmail.com.edu}
{piyanat.kittiwisit@gmail.com}},
Judd D. Bowman$^{2}$, 
Steven G. Murray$^{2}$,
Bharat K. Gehlot$^{3,2}$,
\newauthor{Daniel C. Jacobs$^{2}$ and Adam P. Beardsley$^{4,2}$}
\\\\
$^{1}$Department of Physics and Astronomy, University of the Western Cape, Cape Town, 7535, South Africa \\
$^{2}$School of Earth and Space Exploration, Arizona State University, AZ, USA\\
$^{3}$Kapteyn Astronomical Institute, University of Groningen, PO Box 800, 9700AV Groningen, The Netherlands\\
$^{4}$Physics Department, Winona State University,
Pasteur 120, Winona, MN 55987, USA}
\date{Accepted XXX. Received YYY; in original form ZZZ}

\pubyear{2019}

\begin{document}
\label{firstpage}
\pagerange{\pageref{firstpage}--\pageref{lastpage}}
\maketitle

\begin{abstract}
Measurements of the one-point probability distribution function and higher-order moments (variance, skewness, and kurtosis) of the high-redshift 21 cm fluctuations are among the most direct statistical probes of the non-Gaussian nature of structure formation and evolution during reionization. However, contamination from astrophysical foregrounds and instrument systematics pose significant challenges in measuring these statistics in real observations. In this work, we use forward modelling to investigate the feasibility of measuring 21 cm one-point statistics through a foreground avoidance strategy. Leveraging the characteristic wedge-shape of the foregrounds in $k$-space, we apply a wedge-cut filter that removes the foreground contaminated modes from a mock data set based on the Hydrogen Epoch of Reionization Array (HERA) instrument, and measure the one-point statistics from the image-space representation of the remaining non-contaminated modes. We experiment with varying degrees of wedge-cutting over different frequency bandwidths and find that the centre of the band is the least susceptible to bias from wedge-cutting. Based on this finding, we introduce a rolling filter method that allows reconstruction of an optimal wedge-cut 21~cm intensity map over the full bandwidth using outputs from wedge-cutting over multiple sub-bands. We perform Monte Carlo simulations to show that HERA should be able to measure the rise in skewness and kurtosis near the end of reionization with the rolling wedge-cut method if foreground leakage from the Fourier transform window function can be controlled.

\end{abstract}






\section{Introduction}
\label{sec:intro}

Observations of the high-redshift 21~cm line from neutral hydrogen are direct probes of the intergalactic medium during the first billion years of the universe, and will offer essential information about the astrophysical processes governing structure formation and evolution during the Cosmic Dawn and the Epoch of Reionization (EoR). The recent evidence of detection of the sky-averaged 21 cm spectrum by the Experiment to Detect the Global EoR Signal \citep[EDGES;][]{bowman.etal.2018} and the contrasting results from the Shaped
Antenna measurement of the background RAdio Spectrum experiment 
\citep[SARAS;][]{singh.etal.2022} has marked a new milestone in the study of these cosmological epochs, but complete explanation of the EoR will require studies of the statistical properties of spatial fluctuations in the 21~cm tomography using radio interferometric arrays. 

Ongoing efforts from different radio interferometers over the past decade have yielded several upper limits on the power spectrum of 21~cm fluctuations.\footnote{see recent review by \citet{liu.shaw.2020} for comprehensive introduction to 21~cm data analysis.} These include measurements from the Murchison Widefield Array \citep[MWA;][]{dillon.etal.2014,dillon.etal.2015,beardsley.etal.2016,ewall-wice.etal.2016, barry.etal.2019,trott.etal.2020,rahimi.etal.2021}, Low Frequency Array \citep[LOFAR;][]{patil.etal.2017,gehlot.etal.2019,mertens.etal.2020}, the Giant Metrewave Radio Telescope
\citep[GMRT;][]{paciga.etal.2013}, The Owens Valley Long Wavelength Array \citep[OWRA-LWA;][]{eastwood.etal.2019,garsden.etal.2021}, The Donald C. Backer Precision Array for Probing the Epoch of Reionization \citep[PAPER;][]{parsons.etal.2014,jacobs.etal.2015,kolopanis.etal.2019}, and the partially completed Hydrogen Epoch of Reionization Array \citep[HERA;][]{heracollaboration.etal.2022}. In particular, HERA is nearing completion of continual built-out to its full array with 350 antennas and is expected to be able to make a high sensitivity detection \citep{deboer.etal.2017}.

Despite the promise of these experiments, significant information will be lost in power spectrum analyses due to the non-Gaussian nature of the 21\,cm intensity fluctuations. As reionization progresses, ionised regions will form around groups of radiating sources, causing the distribution of 21~cm intensity field to deviate from the nearly Gaussian underlying matter density field (see, e.g., \citealt{ciardi.madau.2003,furlanetto.etal.2004,bharadwaj.pandey.2005,cooray.2006, iliev.etal.2006,mellema.etal.2006,lidz.etal.2007}). Since a power spectrum is not sensitive to non-Gaussianity, different underlying distributions of the fluctuations could result in the same power spectrum; higher-order statistics will be needed to fully describe the signal.

One promising set of alternative statistics to probe non-Gaussianity in the EoR is the one-point probability distribution function (PDF) of the 21~cm brightness temperature intensity fluctuations and its higher-order moments (e.g., variance, skewness, and kurtosis), often collectively referred to in the literature as 21~cm one-point statistics. The two primary appeals of these statistics are their simplicity and direct sensitivity to non-Gaussianity. Various aspects of 21~cm one-point statistics have been explored in literature over the last two decades. Early works primarily focused on the theoretical aspects of the statistics to establish them as indicators for reionization signals \citep{wyithe.morales.2007,harker.etal.2009,ichikawa.etal.2010,gluscevic.barkana.2010}. Some of the later works have shown that the variance and skewness react to changes in reionization scenarios rather distinctively, suggesting that these statistics can be used to discern different reionization models \citep{watkinson.pritchard.2014,watkinson.pritchard.2015,watkinson.etal.2015,kubota.etal.2016}. In \citet{kittiwisit.etal.2018} (hereafter \citetalias{kittiwisit.etal.2018}), we performed sensitivity analysis of HERA to 21~cm one-point statistics using mock observations that account for systematics from: (a) instrument configuration, (b) thermal noise, (c) sample variance, and (d) lightcone effect \citep{zawada.etal.2014,mondal.etal.2018a}. We focused on the variance, skewness, and kurtosis measurements and experimented with frequency binning and windowing to improve sensitivity. We concluded that the full HERA array will be able to measure 21 cm one-point statistics with sufficient sensitivity, assuming that contamination from astrophysical foreground sources are mitigated. 

However, removal of foreground contamination is extremely challenging in practice due to complications from the interaction between the foregrounds and the frequency-dependent instrument responses known as mode-mixing. In Fourier space, the chromatic response of a multi-baseline interferometer causes the intrinsic foregrounds (confined to the lowest line-of-sight $k_\parallel$ modes) to mix with the higher-$k$ modes. This leads to a wedge-shaped region in ($k_{\perp}, k_{\parallel}$) space, obscuring the 21~cm signal of interest. The foreground wedge has been studied extensively literature, including simulation works that show the effects of the wedge on the 2D \citep{datta.etal.2010,hazelton.etal.2013,thyagarajan.etal.2015a,thyagarajan.etal.2016} and 1D \citep{morales.etal.2012,vedantham.etal.2012,trott.etal.2012} power spectrum space, analytical works that provide rigorous mathetical framework \citep{liu.etal.2014,liu.etal.2014a}, and works that explore effects of the foreground wedge on post-EoR 21~cm signals \citep{seo.hirata.2016}. To model and subtract the foreground wedge from the signal, both the foregrounds and the instrument must be precisely known, a very demanding requirement as observations of the former are lacking in the relevant frequency range and the latter are usually extremely complex. Due to this difficulty, most works on 21~cm one-point statistics, including \citetalias{kittiwisit.etal.2018}, only include simple instrumental effects. One exception is the work by \citet{patil.etal.2014}, in which the authors have utilised the well-developed software infrastructure for foreground modelling and subtraction of the LOFAR telescope into sensitivity analysis of the telescope to the 21~cm variance.

An alternative approach to foreground removal is the so called ``foreground avoidance'' strategy, in which the Fourier modes contaminated by the foregrounds and instrumental response are removed from the analyses \citep{parsons.etal.2012}. The pros and cons of foreground avoidance strategy have also been extensively studied in literature including how it would bias the 21~cm power spectrum \citep{jensen.etal.2016}. Comparison with foreground subtraction strategy has also been studied \citep{chapman.etal.2016}. Recent 21~cm power spectrum results from several 21~cm experiments have utilised foreground avoidance analysis to yield improved upper limits on the 21 cm power spectra \citep{kolopanis.etal.2019,trott.etal.2020,garsden.etal.2021,heracollaboration.etal.2022}. In principle, the foreground avoidance strategy can also be used in other 21 cm statistics as the mathematical descriptions of the foreground wedge in the Fourier Transform space is well defined and independent of the measured quantities. However, discarding the data in the wedge will bias the measurements of the statistics that would otherwise incorporate those modes.

In this paper, we investigate the feasibility of measuring the 21~cm one-point statistics under the foreground avoidance regime through simulation and analyses of a mock data set. We construct foreground wedge filters based on the mathematical description of the foreground wedge, and apply them following a conventional procedure to mock HERA data previously developed in \citetalias{kittiwisit.etal.2018}. To develop insight into the bias from foreground wedge cutting on the measured one-point statistics, we assess the impact of gradually increasing the extent of the wedge-filter, corresponding to a range of cases from pessimistic to optimistic. Based on this finding, we develop a rolling filter method, which we will demonstrate can produce an optimal wedge-cut intensity map. Lastly, we apply the rolling wedge-cut filters on our mock data set and perform Monte Carlo simulations to establish a baseline expectation of the detectability of one-point statistics using foreground wedge-cut filters.

The rest of the paper is organised as follows. In Section~\ref{sec:method}, we describe our mock data set and give an overview of one-point statistics. The variance, skewness, and kurtosis derived from this mock data are presented as references for the rest of the results from this work. Section~\ref{sec:conventional_wc} discusses the analysis steps and results of the conventional foreground wedge-cutting. A brief overview of on the physical and mathematical nature of the foreground wedge is also provided in Section~\ref{sec:foreground_wedge}. Section~\ref{sec:rolling_wc} describes our rolling wedge-cut method, and the details and results from the Monte Carlo simulations. Finally, we conclude in Section~\ref{sec:conclusion}.

\section{Methodology}
\label{sec:method}

\subsection{Mock Data}
\label{sec:data}

All analyses in this work utilise a mock data set developed in \citetalias{kittiwisit.etal.2018}. The model consists of pure 21 cm signals derived from semi-analytic 21~cm simulations provided by \citet{malloy.lidz.2013} with added simple point-spread-function (PSF) effects based on the Hydrogen Epoch of Reionization Array (HERA; \citealt{deboer.etal.2017}) instrument. To summarise, we transformed four-dimensional (three spatial and one time) simulation cubes that span ionised fraction $\sim 0.3-0.96$ into a three-dimensional cube in an observer frame (two angular and one frequency) by using the tile-and-grid method described in Appendix A of \citetalias{kittiwisit.etal.2018}. This observer-frame cube is essentially a set of full-sky 21~cm maps in HEALPix coordinates, with each map corresponding to an 80-kHz frequency channel within the $\sim 139-195$ MHz frequency range that HERA observes. Each map is then smoothed with a Gaussian kernel corresponding to the angular resolution of the HERA instrument with 331 antennas (the core of the full HERA array) at that frequency. This does not add sampling and chromatic effects from the PSF of the telescope, but smooths the intensity fluctuations to the size-scale that can be probed by HERA. Finally, a $\sim 15^{\degr} \times 15^{\degr}$ field, representing the HERA instrument field of view, is randomly picked and then extracted from each map, sine projected, and stacked along the frequency dimension to produce mock observed data cubes. In principle, cutting the image to the field of view does not remove contamination from foreground sources outside of the field interacting with sidelobes of the PSF \citep{pober.etal.2016}, but, because we are not adding real PSF effects as stated above, the field cutting here simply mimics the sizes of the images) that we might be able to obtain from HERA observations. Each cube is $256 \times 256 \times 705$ pixels in the sky positions and frequency. The angular pixel size is $\sim 3.4^{\arcmin}$, oversampling the angular resolution of HERA350 instrument by a factor of $\sim6-8$ in the selected frequency range. 

\subsection{One-point Statistics}
\label{sec:stats_overview}

``One-point statistics'' is often used as a collective term for the probability distribution function (PDF) and moments of a random variable. The PDF is a quantitative measurement of a distribution of a random variable. In the context of 21 cm cosmology observation, this random variable is usually the 21 cm brightness temperature. Then, given a map of 21 cm brightness temperature intensity fluctuations, the PDF is just a histogram calculated from all independent pixels, normalised to an integral of one. It is the simplest statistic that can be used to quantify the distribution of the ionised bubbles and neutral islands in the 21 cm fluctuations across a brightness temperature range, and the degree at which these fluctuations deviate from a Gaussian random field. 

The moments are shape parameters of the PDF. The first moment is the mean (or the expected value) of a distribution, which gives the representative central location of the distribution. The second moment is the variance, which quantifies the spread of the distribution about the mean. The third moment is the skewness, which describes the symmetry (or asymmetry) of the distribution. A negative skewness indicates that the negative tail of the distribution (values that are lower than the mean) is more extended than the positive tail, whereas a positive skewness indicates the opposite. Kurtosis describes the tailedness of the distribution. A positive kurtosis indicates that more mass is concentrated on the tails of the distribution than around its peak, and similarly a negative kurtosis indicates the opposite. A perfect Gaussian distribution can be fully described by just its mean and variance and has zero skewness and kurtosis. Thus, non-zero skewness and kurtosis can provide good checks for non-Gaussianity in reionization.

\begin{figure}
    \centering
    \includegraphics[width=\columnwidth]{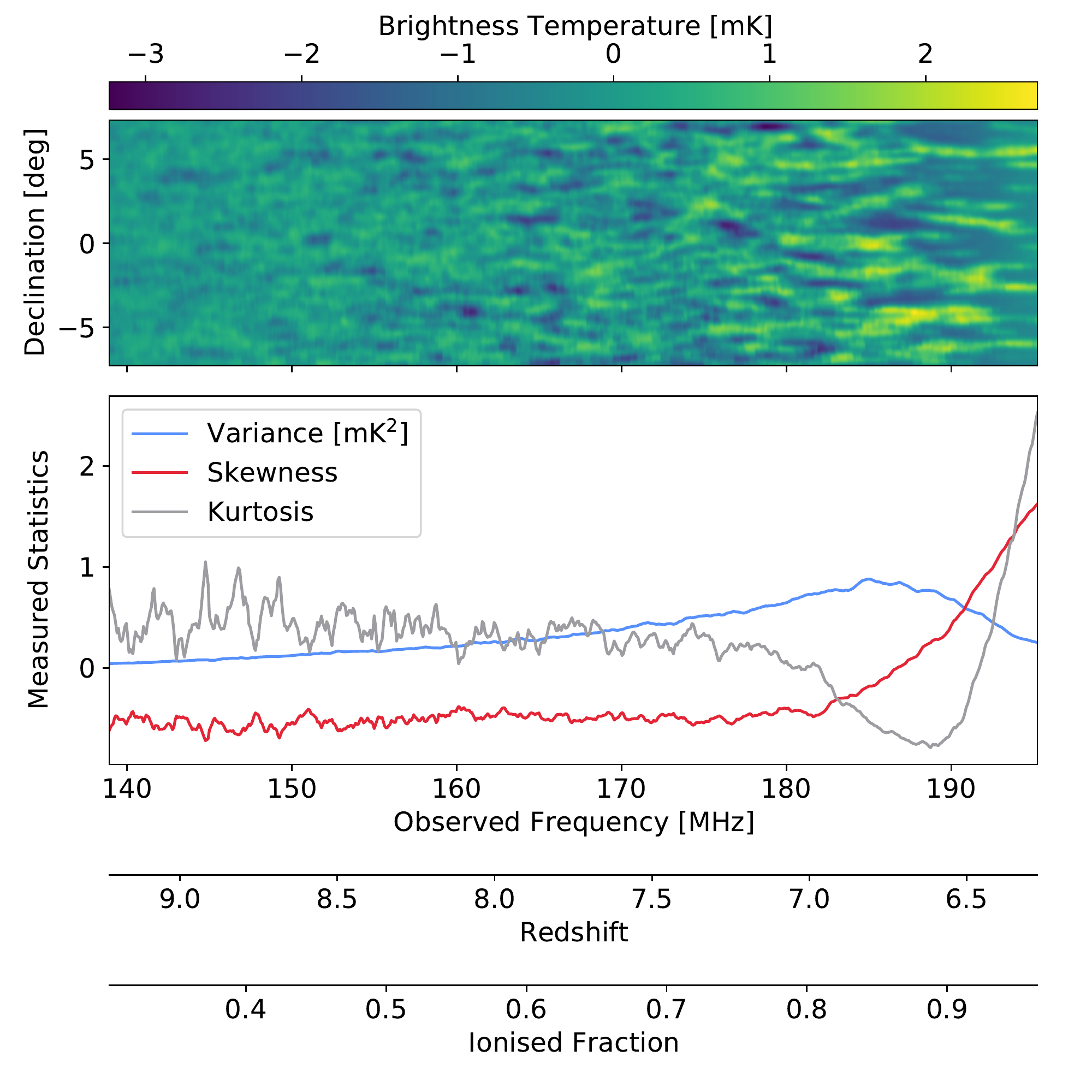}
    \caption{Intensity map (top) and measured one-point statistics (bottom) of the HERA mock data set. We developed this mock data set based on forward-modelling of the HERA instrument in \citetalias{kittiwisit.etal.2018}. No foreground and noise are present in the simulation. The data cube is Gaussian smoothed and mean-subtracted per frequency to mimic interferometric measurements. The intensity map is a lightcone slice at the centre of the cube, showing the evolution of ionised bubbles that formed around neutral islands across a range of observing frequency thought to correspond to the EoR. The variance, skewness, and kurtosis exhibit unique evolution following the evolution of the fluctuation with the prominent rising of skewness and kurtosis near the end of reionization that provide strong signals for targeted observations.}
    \label{fig:model}
\end{figure}

Mathematically, variance, skewness, and kurtosis are derived from central moments as given in probability theory. For $N$ samples of a random variable with values $x_i$, where $i = 1,\dotsc,N$, and a mean value $\overline{x} = \sum_{i=0}^{N} x_i / N$, the $p$-th order central moment $m_p$ is given by,
\begin{gather}
    m_p = \frac{1}{N} \sum_{i=0}^{N} (x_i - \overline{x})^p.
    \label{eq:moments_def}
\end{gather}
Then, the variance $S_2$, skewness $S_3$, and kurtosis $S_3$ are as defined followed,
\begin{align}
    S_2 &= m_2, \label{eq:v_def} \\
    S_3 &= \frac{m_3}{(m_2)^{3/2}}, \label{eq:s_def} \\
    S_4 &= \frac{m_4}{(m_2)^2} - 3. \label{eq:k_def}
\end{align}
We plot the variance, skewness and kurtosis of the mock data cube along with the lightcone cross-section showing the evolution of the 21 cm fluctuations in Figure~\ref{fig:model}. The mock data cube is mean subtracted per frequency before calculating the statistics. These statistics measured from the mock dataset with no foreground wedge cutting will be referred to as reference statistics hereafter.

As discussed in \citetalias{kittiwisit.etal.2018}, the PDF of the 21 cm brightness temperature intensity fluctuations has a unique evolution across redshift, yielding distinct functions of variance, skewness, and kurtosis. In summary, early in reionization, the 21 cm PDF is a near Gaussian distribution with its peak centred at a positive temperature value and a long negative tail that corresponds to cold spots in the 21 cm fluctuations. This yields a negative skewness and a positive kurtosis. As reionization progresses, the PDF transitions into a bi-modal distribution with a second peak forming and growing at zero temperature. This peak corresponds to ionised regions being formed. During this phase, the variance and skewness increase while the kurtosis decreases. The variance reaches a maximum at a certain redshift, which in our model is located at $\sim 185$ MHz, near where the skewness transitions from negative to positive value and the kurtosis approaches its minimum. Eventually, the 21 cm fluctuations become mostly ionised, producing a delta-like PDF with a sharp peak at zero and a long positive tail corresponding to the reminding small neutral islands. At this final phase, the skewness and kurtosis sharply increase while the variance sharply decreases. 

There are two caveats in this picture. First, as a random variable statistic, the PDF is a function of the sampling resolution. This means that, in a theoretical setup, e.g. given a 21 cm simulation cube, the measured PDF depends on the pixel resolution. In a real observation, the angular resolution of the telescope defines this resolution while the limited field of view adds sample variance fluctuations to the statistics. Secondly, because interferometers do not measure the total power but fluctuations, the PDF is expected to be centred around zero regardless of the global 21~cm brightness temperature level. Fortunately, the variance, skewness and kurtosis do not change as they are calculated about the mean of the distribution.

\section{Conventional Foreground Wedge Cut}
\label{sec:conventional_wc}

To investigate foreground avoidance strategy on the measurements of one-point statistics, we will construct foreground wedge-cut filters with various degrees of filtering, apply them to our mock data cube, and calculate and compare variance, skewness, and kurtosis of the resulting data cube. Wedge-cutting is a common method of removing foreground contaminated data from an analysis. It requires knowledge of the shape of the foreground wedge and effects of the Fourier transform window function, which we summarise in Section~\ref{sec:foreground_wedge}. 
Although the measurements made from the wedge-cut data will be biased, questions such as, ``How much of the wedge should be cut?'', ``Over what size of frequency bandpass should the wedge be cut?'', ``Does cutting the wedge on a smaller bandpass yield better measurements?'', must be answered to gauge the feasibility of this particular analysis. We explore these questions in Section~\ref{sec:wc_bias}.

\subsection{The Foreground Wedge}
\label{sec:foreground_wedge}

We adapt a mathematical description of the foreground wedge from \citet{morales.etal.2012}. To summarise, due to mode-mixing, contamination from a foreground source with a smooth spectrum transiting at an angle $\theta$ from the centre of the observing field will appear in the 2D, line-of-sight and spatial, Fourier transform space ($k_{\parallel}$, $k_{\perp}$) as a line that follows a linear relationship,
\begin{gather}
    k_{\parallel} = \sin(\theta)\frac{H_0 E(z) D_M(z)}{c (1 + z)}k_{\perp},
    \label{eq:ch3_wedge_boundary}
\end{gather}
given the Hubble constant $H_0$, the transverse comoving distance $D_{M}(z)$, and $E(z) = [\Omega_M (1 + z)^3 + \Omega_k (1 + z)^2 + \Omega_{\Lambda}]^{1/2}$ describing the assumed cosmology at a reference redshift $z$. 

\begin{figure}
    \centering
    \includegraphics[width=\columnwidth]{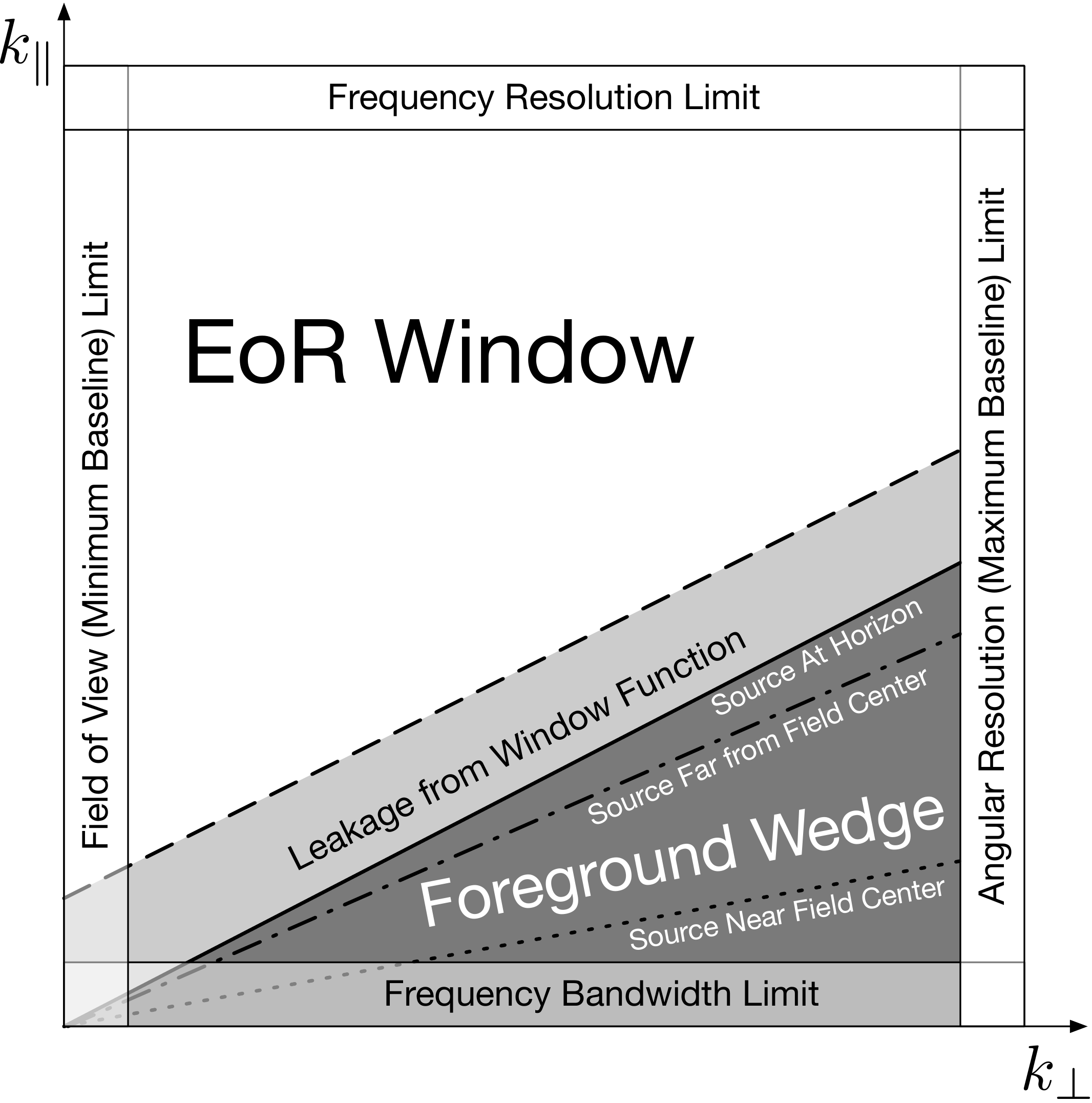}
    \caption{Schematic diagram of the foreground wedge. Chromaticity in multi-baseline interferometer causes power of foreground sources at different angle distances from the field centre to appear as lines in ($k_{\parallel}, k_{\perp}$) with shallower and steeper slope angles for sources near the field centre and far from the field centre respectively. This so called mode-mixing contamination from multiple sources across the field accumulates to occupy a wedge-shape region in ($k_{\parallel}, k_{\perp}$). The Fourier transform window function will cause the foreground to bleed outside of the wedge to the higher $k_{\parallel}$ modes.}
    \label{fig:ch3_wedge}
\end{figure}

Figure~\ref{fig:ch3_wedge} illustrates the consequences of this relationship. A foreground source near the field centre will appear as a line of contamination with a shallow slope angle in the $k_{\parallel}$--$k_{\perp}$ space (dotted line), whereas a source closer to the horizon will appear with a steeper slope (dash-dotted line). Hence, multiple lines of contamination with different slopes from multiple sources across the observing field will form the foreground wedge (dark gray shade). Sources at the horizon, $\theta=\pi/2$ radians, define the \emph{geometric horizon} boundary of the wedge (solid line). The measurable ($k_{\perp}$, $k_{\parallel}$) are limited by observing parameters as annotated in the figure. Note that we plot Figure~\ref{fig:ch3_wedge} in linear ($k_{\parallel}$, $k_{\perp}$) while often the 21 cm power spectrum is plotted in a log scale.

In addition to the mode-mixing effects, a Fourier transform operation with finite boundaries will produce spectral leakage that spill foreground power from the wedge to the EoR window. The spectral structure of this leakage depends on the window function being used. Blackman-Harris \citep{blackman.tukey.1958} or Blackman-Nuttall \citep{nuttall.1981} windows are popular choices in power spectrum analysis due to their high dynamic range, i.e. their Fourier transforms have very low power in the sidelobes relative to the peaks \citep{vedantham.etal.2012, thyagarajan.etal.2013}. However, the mainlobes of their spectra are also broad (as opposed to a rectangular window, i.e. no window function). This renders the $k_{\parallel}$ modes neighbouring the geometric horizon boundary of the foreground wedge unusable, which we show as a light grey region with a dashed-line boundary in Figure~\ref{fig:ch3_wedge}. We will account for this smearing by shifting the boundary of the wedge cutting to higher $k_{\parallel}$ beyond the geometric horizon boundary. This shifting is equivalent to adjusting the $k_{\parallel}$ intercept in Equation~\ref{eq:ch3_wedge_boundary}, which we indicates with a variable $\beta$ in a modified expression for the wedge boundary below,
\begin{gather}
    k_z = \sin(\theta)\frac{H_0 E(z) D_M(z)}{c (1 + z)}\sqrt{k_x^2 + k_y^2} + \beta.
    \label{eq:ch3_wedge_bound_3d}
\end{gather}
Here, we also substitute $k_z = k_{\parallel}$ and $k_{\perp} = \sqrt{k_x^2 + k_y^2}$ to represent the relationship in the 3D Fourier transform space $(k_x, k_y, k_z)$, where we adopt the mapping between the observing spatial and frequency coordinates to $k$ following the mathematical framework for coordinate transformation between visibility and power spectrum space in \citet{morales.hewitt.2004}. In this 3D space, the mode-mixing foreground contamination appears as two concentric cones with their tips touching at $(k_x,k_y,k_z) = 0$. The units of $\beta$ are in pixels of $k_z$, which varies depending on the frequency bandwidth of the data cube. In practice, the value of $\beta$ depends on (and is often fixed by) the choice of window function. For example, the Fourier spectrum of a Blackman-Nuttall window drops by -60 dB at 4 $k$ bins from its peak, and thus a four-pixels buffer ($\beta=4$) is required to suppress foregrounds by 60\,dB. Often times more $k$ bins are removed beyond this number to provision against additional foreground leakage. However, cutting more $k$ modes will remove more signal; thereby decreasing performance (as we will show). 

The EoR window outside of the foreground wedge is in principle free from mode-mixing contamination, although calibration errors, variation in the antenna beams or positions, and other systematics can bleed the foreground power to the outside of the wedge (see, e.g., \citealt{byrne.etal.2019,orosz.etal.2019,joseph.etal.2020,choudhuri.etal.2021}). Hypothetically, if we could control the extent of the foreground sources to be within a certain angle $\theta$ from the field centre, we could have a smaller foreground wedge and bigger EoR window to work with, and thus less contaminated measurements. This is practically impossible, but there is a possibility that the wedge angle or the power inside the wedge could be reduced as instrument knowledge improves (see, e.g., \citealt{murray.trott.2018}). Hence, we will experiment with different $\theta$, for it could also give information on which modes in the $k$ space most influence the measured one-point statistics.

\subsection{Bias from Foreground Wedge Cutting}
\label{sec:wc_bias} 
We conduct three tests to investigate how different wedge-cutting will affect the measured one-point statistics. For the first test, we investigate the effect of the bandwidth of the wedge-cut filters by performing wedge-cutting on subbands of the mock dataset with bandwidths $\wcbw=$ 1, 2, 4, 6, and 8 MHz. We keep the degree of filtering minimal for this test by setting the $\theta$ and $\beta$ parameters in Equation~\ref{eq:ch3_wedge_bound_3d} to 5 degree and 0 pixel when constructing the wedge-cut filters. This roughly corresponds to assuming that foregrounds only exist inside the field of view of the HERA instrument and no leakage occurred from the Fourier transform. For the second test, we experiment with different degrees of filtering by setting $\theta$ to 5, 15, 30, 60, and 90 degrees while keeping $\wcbw=8$ MHz and $\beta=0$ pixel. For the third test, we focus on the outcome of extending the wedge boundary in $k_{\parallel}$ to account for leakage of the Fourier transform when doing a wedge-cut by varying $\beta$ from 0 to 4 pixels while keeping $\wcbw=8$ and $\theta=5$ degree. All choices for the values of $\theta$ and $\beta$ other than the combination of $\theta=90^\circ$ and $\beta>0$ are arguably unrealistic. As discussed, the choice for $\beta$ in practice is fixed by the type of the window function, and there is not yet a possible way to limit the extent of the foregrounds to allow $\theta < 90^\circ$, but we have allowed some flexibility on the filter parameters here to disentangle them from one another and to make it easier to interpret the results. Because the period near the end of reionization that yields the variance peak and the strong rising in skewness and kurtosis, which are the best targets for detection, only occupy the last 10~MHz of our model (see Figure~\ref{fig:model}), we only test up to the maximum subband bandwidths of 8 MHz. This also allows us to centre all subbands in the test (regardless of their bandwidths) at $\sim 191$ MHz.

The analysis steps are the same for all tests except for the parameters of the wedge-cut filters as specified. First, we select a subband of the mock dataset centred at a chosen frequency, multiply the resulting data cube with a Blackman-Nuttall window of the same $\wcbw$ bandwidth along the frequency dimension to shape the bandpass, and Fourier transform to $k$ space. Then, we construct a foreground wedge-cut filter in $k$ space with zero and one for the modes to cut and to keep, following the geometry described in Equation~\ref{eq:ch3_wedge_bound_3d} and making sure that all pixels that are partially inside the wedge are also cut. Thus, the filter bandwidth $\wcbw$ defines both the bandwidth of the data cube and the bandwidth window function, as well as the $k$ space resolution of the filter. We use the redshift corresponding to the mean frequency of the subband as the reference redshift $z$  in Equation~\ref{eq:ch3_wedge_bound_3d} when constructing the filter. All other cosmological parameters are taken from the 2015 Planck results \citep{planckcollaboration.etal.2016}. Next, we multiply the wedge-cut filter with the Fourier representation of the subband cube to remove foreground contaminated modes, and invert Fourier transform back to spatial--frequency space. Finally, we divide the resulting cube with the same Blackman-Nuttall window used during the forward transform to undo the window function, and calculate the statistics.

\begin{figure*}
    \centering
    \includegraphics[width=0.85\textwidth]{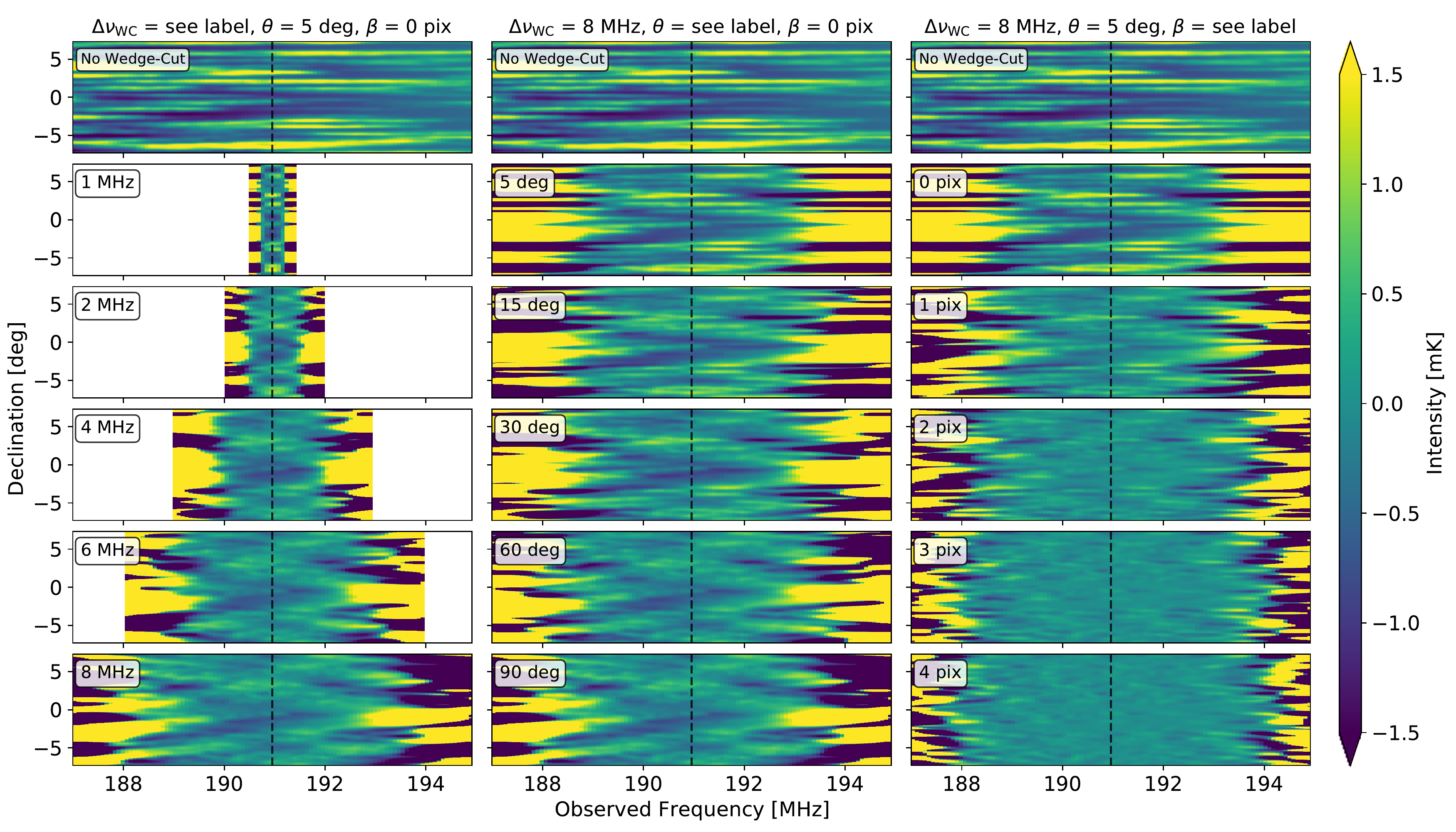}
    \caption{Frequency slices of the intensity maps after foreground wedge-cutting for our three tests. The first test on the left column experiments with wedge-cutting over different sizes of frequency subbands centred at $\sim 191$ MHz. The second test in the middle column shows effects from wedge-cutting assuming different wedge-angle. The third test performs a minimal wedge-cut (assume shallow wedge angle) but shift the wedge boundary to higher $k$ by the numbers of pixels as labelled. Notice the saturation at the edges of the subband due to window function un-weighting, rendering the affected portions unusable for measurements. In the middle of the bands, the foreground wedge-cut maps appear to have more reduced dynamic range and washed out structures as more modes are removed. Comparing the second and the third test shows that shifting the wedge-boundary to higher $k$ is more costly than using steeper wedge angle, which suggests that removing the large scale modes tends to decrease non-Gaussianity.}
    \label{fig:wc_maps_lightcone}
\end{figure*}

\begin{figure}
    \centering
    \includegraphics[width=0.98\columnwidth]{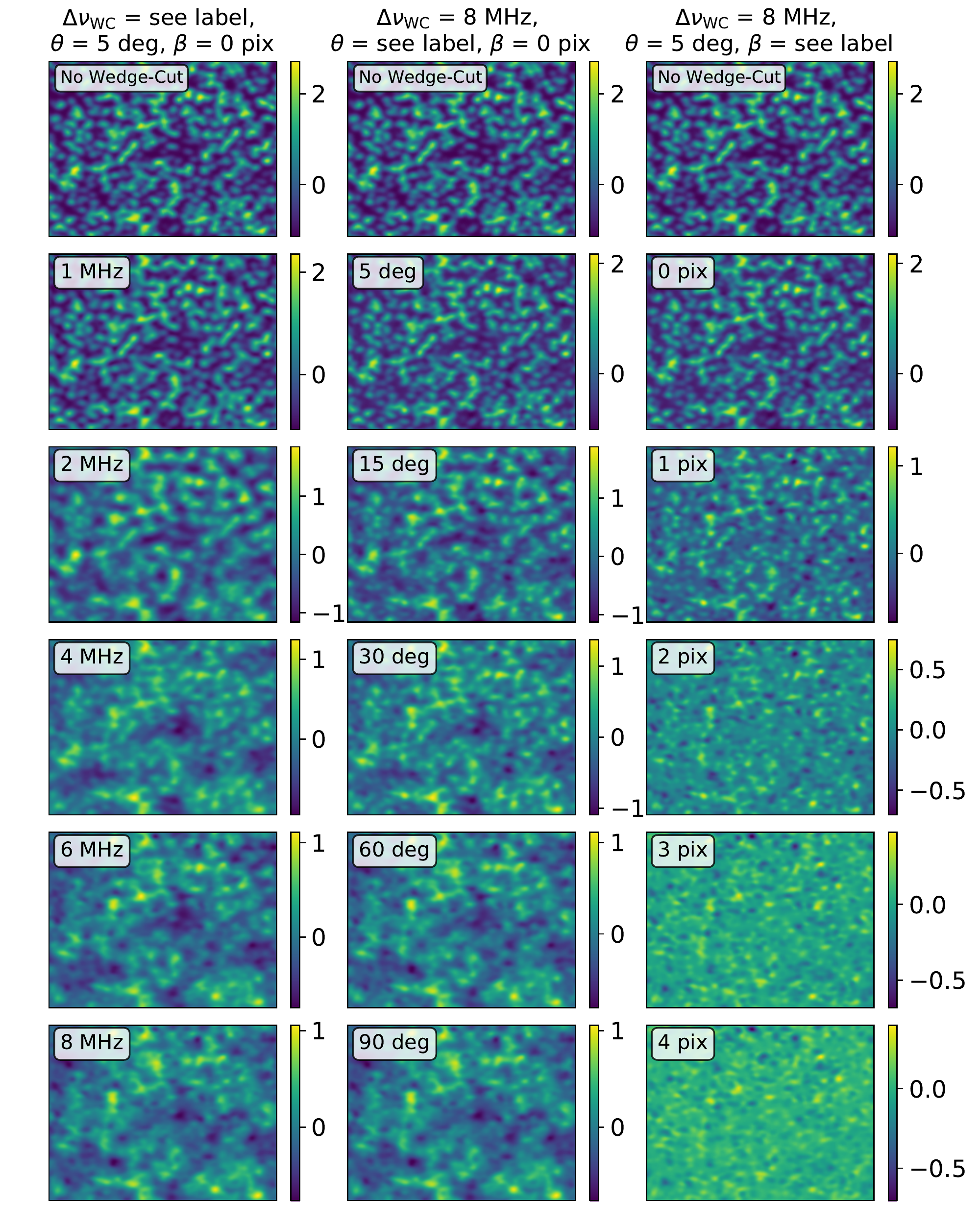}
    \caption{Transverse slices of the the intensity maps at the middle of the band (190.96 MHz) after foreground wedge-cutting. Each panel corresponds to the same test case as in Figure~\ref{fig:wc_maps_lightcone}.}
    \label{fig:wc_maps_transverse}
\end{figure}

\begin{figure*}
    \centering
    \includegraphics[width=0.9\textwidth]{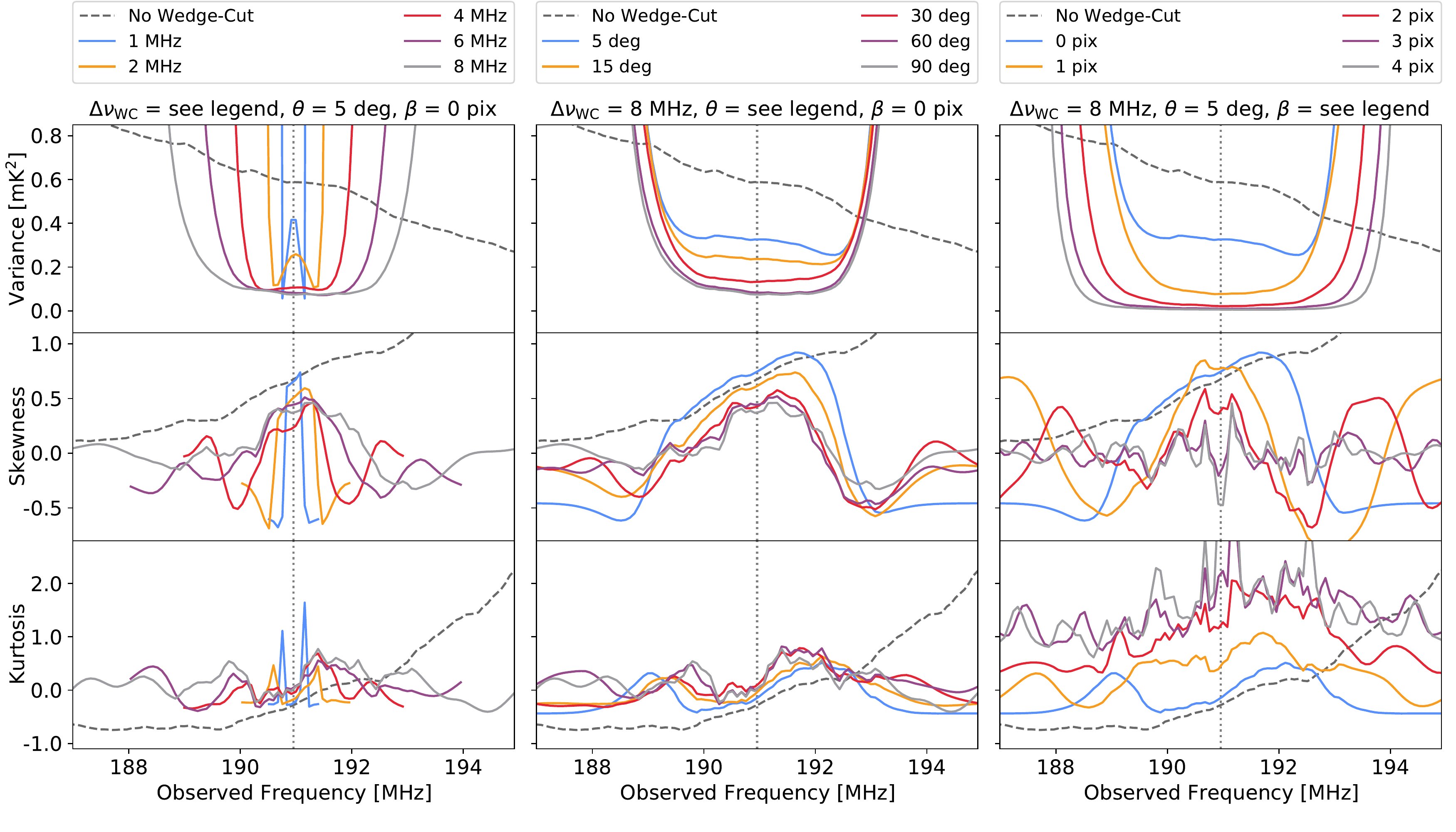}
    \caption{One-point statistics measured from wedge-cut intensity maps for our three test cases. The overshoot and saturation at the edges of the subbands are due to the window function un-weighting and should be discarded from actual measurements. Statistics measured from the middle of the subbands are visibly biased from wedge-cutting. As more $k$ modes are cut, the variance reduces to near zero, and the skewness and kurtosis become noisy as well as reducing to near zero. These indicate that wedge-cutting removes non-Gaussian structures from the intensity maps, which is apparent in Figure~\ref{fig:wc_maps_lightcone}.}
    \label{fig:wc_stats}
\end{figure*}

\begin{figure*}
\centering
    \includegraphics[width=0.85\textwidth]{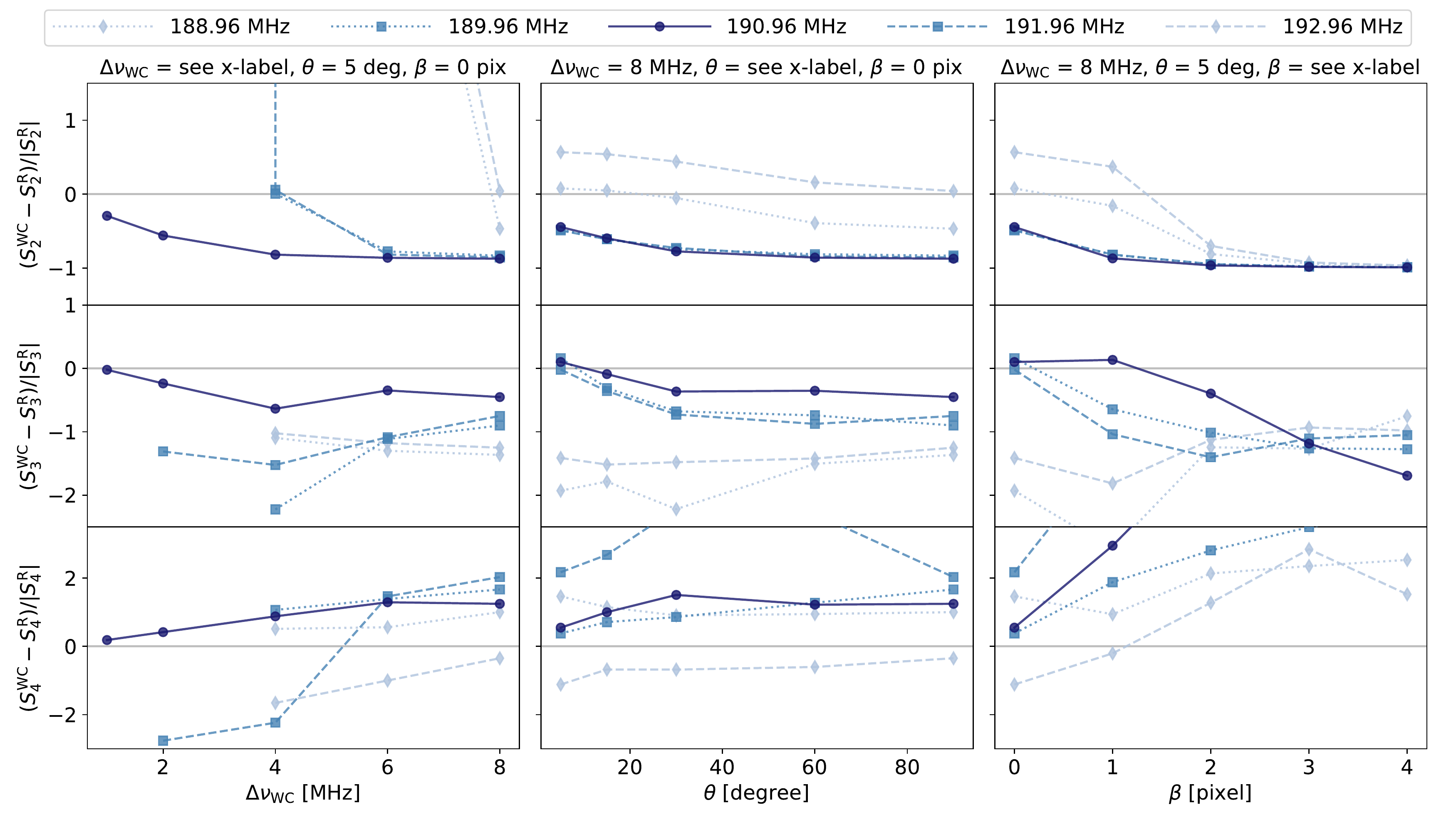}
    \caption{Fractional deviation due to wedge-cutting for our three test cases. We define the fractional deviation due to wedge-cutting as, $(S_{i}^\mathrm{WC} - S_{i}^\mathrm{M}) / |S_{i}^\mathrm{M}|$, where $S_{i}^\mathrm{WC}$ and $S_{i}^\mathrm{M}$ are statistics from the wedge-cut data and the mock data with no wedge-cut respectively. The values plotted here are derived from the measurements in Figure~\ref{fig:wc_stats} at the centre of the band (circle), $\pm1$ MHz (square), and $\pm2$ MHz (diamond). Dashed lines and dotted lines indicate frequency above and below the band centre. In all cases, the centre channel of the subband is the least susceptible to changes in wedge-cut filter parameters. For skewness, the centre channel of the subband are also least biased from wedge-cutting for most cases.}
    \label{fig:wc_stats_3chs}
\end{figure*}

\begin{figure*}
    \centering
    \includegraphics[width=\textwidth]{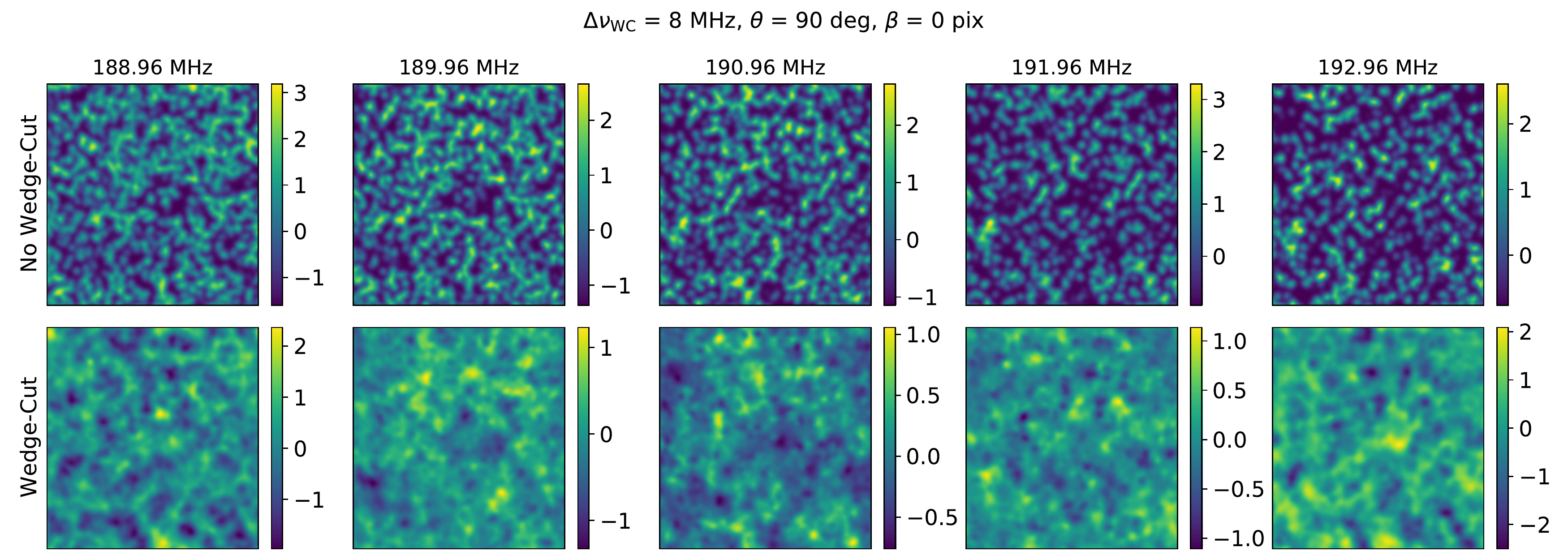}
    \caption{Transverse slices of intensity maps.  The top row shows the reference case with no wedge-cut and the bottom row shows the case of wedge-cut intensity maps for the $\theta=90^\circ$ and $\beta>0$.  The middle column is for the centre channel of the 8~MHz band centred at 190.96~MHz.  The centre channel is the least susceptible in the selected band to bias from wedge-cutting.  By-eye inspection shows the top and bottom panels look more similar at 190.96~MHz than at the other neighbouring frequencies shown on the left and right.}
    \label{fig:wc_maps_transverse_theta90_5chs}
\end{figure*}

Figure~\ref{fig:wc_maps_lightcone} shows frequency slices of the filtered intensity maps from the three tests (left to right columns) while Figure~\ref{fig:wc_maps_transverse} shows the transverse slices at the centre of the bands (190.96 MHz) of all test cases in the same arrangement. One-point statistics measured from these maps are plotted in Figure~\ref{fig:wc_stats}, showing the variance. skewness, and kurtosis in the top, middle, and bottom rows respectively. In addition, we define the fractional deviation due to wedge-cutting as, $(S_{i}^\mathrm{WC} - S_{i}^\mathrm{M}) / |S_{i}^\mathrm{M}|$, where $S_{i}^\mathrm{WC}$ and $S_{i}^\mathrm{M}$ are statistics from the wedge-cut data and the mock data with no wedge-cut respectively. This quantity is zero if the measurements are not biased by wedge-cutting. We plot the deviations at the centre of the frequency band at $\sim191$ MHz, $\sim191\pm1$ MHz, and $\sim191\pm2$ MHz in Figure~\ref{fig:wc_stats_3chs}, and, to accompany this figure, we also plot transverse slices of the $\theta=90^\circ$ and $\beta>0$ case at these frequencies in Figure~\ref{fig:wc_maps_transverse_theta90_5chs}

The first clear effect from wedge-cut filtering is that the signal at the edge of the bandpass is saturated due to down-weighting with the window function. The saturation is obvious in the frequency slices of the intensity maps, which renders approximately half of the band along the band edges unusable and results in overshoots in the measured statistics. In the middle of the band, the intensity maps from the wedge-cut data more closely resemble those from the data with no wedge-cutting, but differences are also clearly visible with a clear trend that the more aggressive the wedge-cutting, i.e., going down from the top panel in the middle column of Figure~\ref{fig:wc_maps_lightcone} and \ref{fig:wc_maps_transverse}, the more structures are being washed out. This is problematic because, as discussed in \citetalias{kittiwisit.etal.2018}, the features in one-point statistics during the later phase of reionization correspond to ionised ``bubbles'' that appear as streaks along the frequency dimension in the observed lightcone intensity maps. Thus, washed out intensity maps will weaken these features in the statistical measurements. True to this statement, the measured variance reduces into an almost flat signal while skewness and kurtosis become noisy and move closer toward zero as the filtering becomes more aggressive. The fractional deviation in Figure~\ref{fig:wc_stats_3chs} also paints a similar picture, in which more aggressive filtering results in statistics that are more biased. 

Interestingly, it is apparent that statistics measured from the central frequency of the band are, in most cases, less susceptible to bias from wedge-cutting than those measured from other frequencies away from the band centre. Figure~\ref{fig:wc_stats_3chs} illustrates this point, by which the fractional deviations due to wedge-cutting at the band centre (darkest solid line with circle marker) are closest to zero for most filter parameter combinations. The transverse slices of the wedge-cut intensity maps in Figure~\ref{fig:wc_maps_transverse_theta90_5chs} indeed show that the central frequency slice preserve more the ionised bubbles present in the reference maps. This may be partly because  the weight of the window function being closest to 1 at the band centre, and partly because the centre of the band has the least mixing from other frequencies when thinking of the wedge-cut filter in the Fourier space as a convolution kernel in the real space, but it is difficult to test this assumption due to how the Fourier modes map to real modes. In any case, the fact that we see the central frequency channel less bias from wedge-cutting is interesting, and we will revisit this point in Section \ref{sec:rolling_wc}.

Going over each of the filter parameters one-by-one, first we find that statistics from the wedge-cut data deviate more from the reference as the filter bandwidth increases, as shown in the left column of Figure~\ref{fig:wc_stats_3chs}. Interestingly, this deviation asymptotes at 8-MHz bandwidth, which we think may be due to strong non-Gaussian features in the statistical signals at higher frequencies getting mixed into lower frequencies and evening out detrimental effects from wedge-cutting. However, the results here must be taken with caution. They do not necessarily suggest that doing wedge-cuts on smaller subbands are better. First, the 21 cm signal is rapidly evolving during this last phase of reionization, and it is difficult to remove signal-dependency from our analysis given the one mock dataset that we have. Secondly, using a smaller bandwidth will yield a window function spectrum that extends to higher $k_{\parallel}$  as well as a coarser $k_{\parallel}$ pixel resolution (as opposed to using a larger bandwidth), meaning that we will have to cut to higher $k_{\parallel}$ to account for the leakage. Since our simulation only has the 21~cm signal and no foreground, the window function simply throws the signal to higher $k_{\parallel}$ when performing wedge-cutting on a smaller bandwidth. Therefore, the wedge-cut maps appear to retain more non-Gaussian structures, and statistics that apparently look less biased. If foregrounds were present in our simulation, the window function would similarly throw foreground power to higher $k_{\parallel}$, and looking at how this mix of foreground and signal leakage affects the statistics is the only clear way to quantify the effect of varying bandwidth. Since it is difficult to determine the single best bandwidth that gives the least biased one-point statistics, we have chosen to use an 8-MHz bandwidth for the rest of our analysis to allow the maximum number of $k$ modes to be kept after wedge-cutting.

Effects from the other two filter parameters -- the maximum angle of foreground sources from the field centre $\theta$, and the extra shift in wedge boundary $\beta$ -- are more clear. In general, removing more modes tends to remove more of the non-Gaussian features in the intensity maps and measured statistics. With the more aggressive filtering, the variance flattens across the band, while skewness and kurtosis appear as random fluctuations near zero. Shifting the wedge boundary to higher $k_{\parallel}$ also appears to be much more harmful to non-Gaussian features than increasing the angle of the wedge. This suggests that removing more large scale modes is more detrimental to the detection of non-Gaussianity, but it is inconclusive and difficult to link non-Gaussianity to certain $k$ modes because, as discussed in studies such as \citet{gorce.pritchard.2019,thyagarajan.carilli.2020} and \citet{thyagarajan.etal.2020}, non-Gaussianity arises from correlation between different Fourier modes.

Regardless, the results here suggest that avoiding the foreground via wedge-cut filtering will yield strong bias in one-point statistics unless the minimal filtering can be done, which will only be possible if we can somehow reduce the foreground wedge to the $k$ modes below the horizon limit and minimising spectral leakage from the window function.

\section{Rolling Foreground Wedge Cut}
\label{sec:rolling_wc}
Although our attempts to recover 21 cm one-point statistics by way of foreground wedge-cutting show clear biases, we notice that the central frequency of the band yields statistics that are less susceptible to biases from wedge-cutting, suggesting that we might be able to improve the measurements by only using the central frequency channel of the subband. To test this idea, we develop and experiment with a rolling filter method, which is done as follows. First, the wedge-cut filter is constructed and applied to a chosen subband as discussed in Section~\ref{sec:conventional_wc}, but only the centre frequency channel of the subband is kept after cutting the wedge. Next, the filter subband is shifted by some numbers of frequency channels, and the first step is repeated, essentially rolling the filter across the bandwidth of the data cube. Finally, all of the saved outputs are stitched together to form a wedge-cut 21 cm intensity cube, and the measurements are made from this cube. This procedure is similar to an overlap-save method \citep{hanumantharaju.etal.2007} commonly used in signal processing to reconstruct a convolution of a long time-domain signal from a series of convolutions of overlapping short intervals of that signal, by which the convolved outputs that are contaminated by aliasing are discarded during the signal reconstruction. Thus, the rolling method simply selects the least biased frequency from data but does not reduce the bias beyond the possible outcome from the wedge-cut filter. This specific implementation to 21~cm maps was first described in \citet{kittiwisit.2019} and recently adopted in \citet{prelogovic.etal.2022} for constructing a corrupted training data set for a machine learning study of 21~cm signals.

\begin{figure*}
    \centering
    \includegraphics[width=0.9\textwidth]{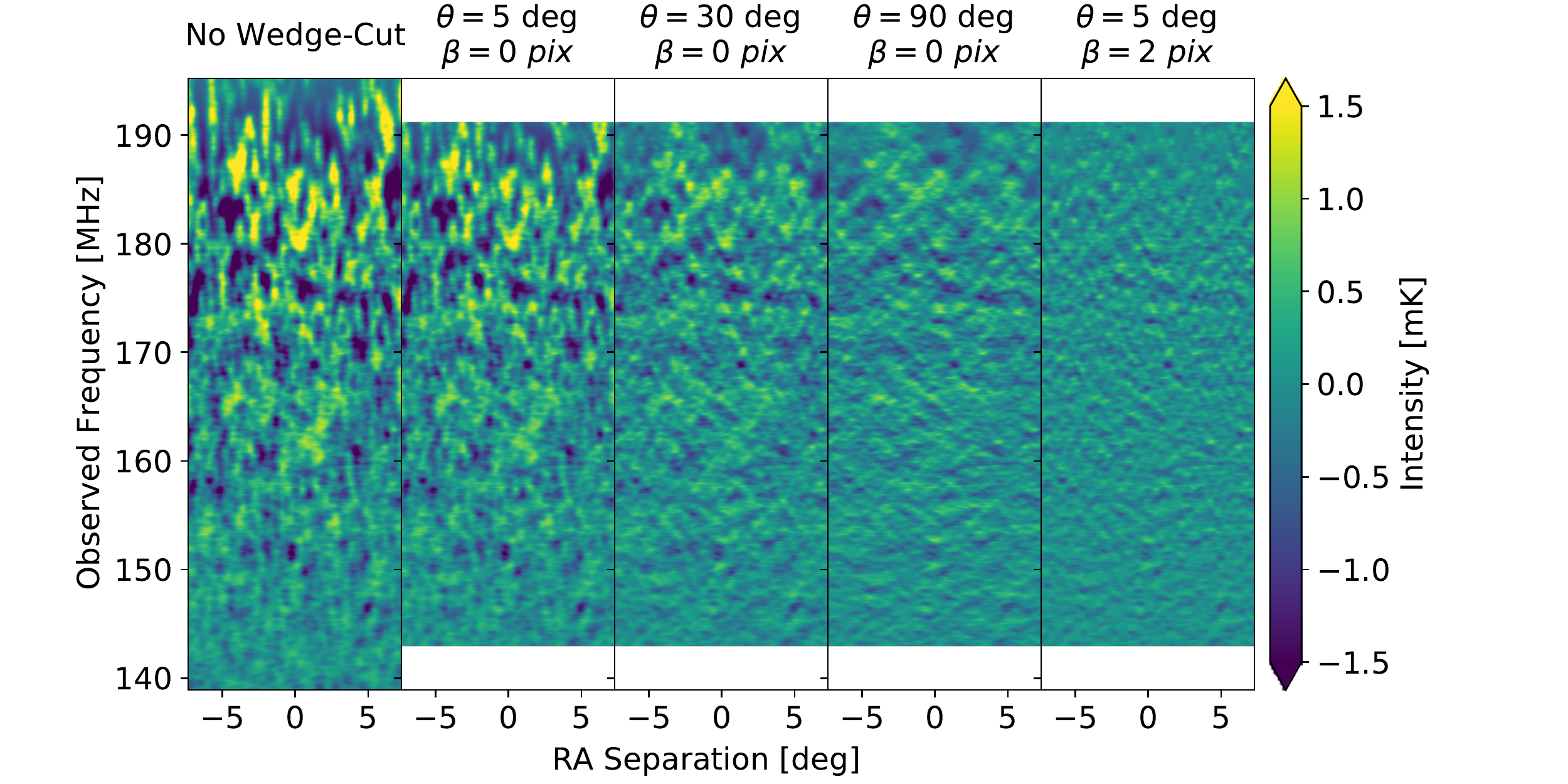}
    \caption{Comparing 21~cm intensity maps after rolling foreground wedge-cutting. Different columns from the left to the right correspond to different degree of wedge-cutting with no-wedge-cut map on the left most column for comparison. The rolling wedge-cut does not reduce bias due to the removal of the wedge, but does allow recovering the full bandwidth of the intensity maps after wedge-cutting. Once again, shifting the wedge boundary to higher $k$ and cutting out more of the lower $k$ modes is more detrimental to the non-Gaussian structures in 21~cm intensity fluctuations. The second and third panels correspond to less aggressive wedge-cutting under assumptions that foreground sources lie within 5 and 30 degree from the field centre, which are unreal scenarios but we think offer good insights into the effect of the foreground wedge-cutting. The fourth panel corresponds to full wedge-cut without accounting for extra leakage, and is the most likely scenario for real observations.}
    \label{fig:rolling_wc_maps_lightcone}
\end{figure*}

Figure~\ref{fig:rolling_wc_maps_lightcone} shows lightcone slices of the mock data before and after a rolling wedge-cut, using an 8-MHz filter bandwidth with four chosen sets of filter parameters that correspond to increasing degrees of filtering from the left to the right panels. These filter parameters also roughly correspond to specific observational scenarios. Counting from the left, the second and third panels use filters with $\theta=5$ and 30 respectively, which are approximately equivalent to assuming that foreground sources only lie within the FWHM and the first null of the primary beam of HERA. The fourth panel uses $\theta=90$, making no assumption on the extent of the foreground sources and cutting them from horizon to horizon. In real observations, foreground sources exist across the sky, but their fluxes are down-weighted by the antenna beam. Thus, none of the cases here exactly represent the real sky, but we find that they offer insightful incremental pictures into the effects of foreground wedge cutting.  As discussed in Section~\ref{sec:wc_bias}, accounting for leakage from the Fourier transform window by setting $\beta>0$ to shift the wedge boundary to higher $k_{\parallel}$ is very costly. To reiterate this point, we show the result from a filter with $\theta=5$ and $\beta=2$ for the last panel, which assumes the minimal extent of the foreground as in the second panel but shifts the wedge boundary to the higher $k_{\parallel}$ by two pixels. This is once again an unrealistic case, but it is very clear that the intensity map here becomes more washed out than one in the fourth panel that uses a full horizon-to-horizon wedge-cut. A close comparison of the reconstructed intensity maps from the rolling wedge-cut to the equivalent maps from our results in Section~\ref{sec:conventional_wc} shows no obvious visual differences apart from the oversaturated band edges in the latter case. This is as expected because the rolling filter does not reduce the bias from wedge-cutting. However, an expanded, almost full bandwidth of usable frequencies is a clear benefit from the rolling method and will make it possible to look at how the unique features in the statistics are affected by wedge-cutting, which we will discuss shortly.

To develop some baseline expectations for HERA of the one-point statistics derived from the rolling wedge-cut intensity maps, we perform  Monte Carlo simulations by generating Gaussian noise cubes and adding them to the mock data cube before performing rolling wedge-cutting with the same set of filter parameters used in Figure~\ref{fig:rolling_wc_maps_lightcone}. We use the Gaussian estimate for the thermal noise RMS in interferometric 21~cm observations given by \citet{watkinson.pritchard.2014} to set the RMS of the noise cubes,
\begin{gather}\label{eq:noise1}
\begin{split}
    \sigma_n = 2.9\,\mathrm{mK} 
    &\left(\frac{10^5\,\mathrm{m}^2}{A_{\mathrm{tot}}}\right)
    \left(\frac{10\,\mathrm{arcmin}}{\Delta\theta}\right)^2 \\
    &\times \left(\frac{1+z}{10.0}\right)^{4.6}
    \sqrt{\left( \frac{1\,\mathrm{MHz}}{\Delta\nu} \frac{100\mathrm{h}}{t_{\mathrm{int}}}\right)}.
\end{split}
\end{gather} 
The specific exponent in the redshift term here comes from an assumption that the system temperature of an interferometer is dominated by the sky brightness, which follows a power law with $-2.6$ exponent (see e.g. Section 9 of
\citet{furlanetto.etal.2006} for a full derivation). Given the HERA instrument with the total collecting area of the array $A_{\mathrm{tot}}=50953$~m$^{2}$ (for the 331-antennas core that will be used in EoR observation), the maximum baseline $D_{\mathrm{max}}=294$~m that can be translated to $\Delta \theta = \lambda / D_{\mathrm{max}}$, and assuming the base integration time $t_{\mathrm{int}}=100$ h and frequency channel $\Delta \nu = 0.08$ kHz matching our data cube, the expression reduces to,
\begin{gather}\label{eq:noise2}
    \sigma_n = 2956.972\ \mathrm{mK} \left( \frac{1\ \mathrm{MHz}}{\nu_{\mathrm{observed}}} \right)^{2.6} .
\end{gather}

We perform three simulations with 1.0, 0.6 and 0.3 times the RMS noise factor, corresponding to 100, 277, and 1111 hours of integration time per field on HERA. Although these fractions were chosen primarily for convenience, they roughly translate to 0.4, 1.1, and 4.5 years of observations assuming $\sim 68\%$ of the observing time are usable (based on the recent internal data release{\footnote{\url{http://reionization.org/manual_uploads/HERA097_H1C_IDR3_2_Memo.pdf}}}). Because the mock data has already been smoothed with a Gaussian kernel matching angular resolution of the telescope per frequency to approximate the instrument PSF effect, the noise cubes are first made and smoothed with the same Gaussian kernels. Then, we re-normalise the noise power to achieve the RMS per frequency that matches Equation~\ref{eq:noise2} before adding to the mock data cube for wedge-cutting. 

As the output intensity maps now contain noise, simply using Equation~\ref{eq:moments_def} to \ref{eq:k_def} to calculate the statistics will result in noise bias. Unbiased estimators for the moments must be used. We have derived them in the Appendix B in \citetalias{kittiwisit.etal.2018}, which are summarised below.
\begin{align}
    \widehat{m}_2 &= m_2^{\mathrm{N}} - \sigma_n^2, 
    \label{eq:ch3_unbiased_m2} \\
    \widehat{m}_3 &= m_3^{\mathrm{N}}, 
    \label{eq:ch3_unbiased_m3} \\
    \widehat{m}_4 &= m_4^{\mathrm{N}} - 6 m_2^{\mathrm{T}} \sigma_n^2 - 3 \sigma_n^4.
    \label{eq:ch3_unbiased_m4}
\end{align}
Here, the unbiased quantities are denoted with the wedge $\wedge$ symbols, and the noise-biased and the \emph{true} (derived from noiseless data) quantities are denoted with superscripts $\mathrm{N}$ and $\mathrm{T}$ respectively. Notice that the 3rd moment is not biased by noise, and the noise variance must be known to obtain the unbiased second and fourth moments. The true second moment is also needed to calculate the unbiased fourth moment, but this is impossible to obtain in real observation, so we use the unbiased second moment as an estimate for the true second moment. To obtain the noise variance, we apply the same set of filters to the noise cubes and measure the variance of the output. In actual observations, the noise variance can be computed by subtracting two subsequent integrations. With the moments derived from unbiased estimators, the variance, skewness and kurtosis can be calculated in the usual way following Equation~\ref{eq:v_def} to \ref{eq:k_def}. Then, the mean and the standard deviation of the measured statistics over the noise realisations are taken as the expectation values and uncertainties. 

\citetalias{kittiwisit.etal.2018} suggested that some forms of frequency averaging will be needed to improve SNR in the measurements of one-point statistics. Thus, we experiment with frequency binning by averaging the noise-added intensity maps over 1 to 8 MHz bandwidths before calculating the statistics. Also following \citetalias{kittiwisit.etal.2018}, we account for sample variance in the drift-scan observations of HERA by repeating the simulation over 19 more observing fields. This was done by picking different $15^{\degr}\times15^{\degr}$ fields from the full-sky HEALPix maps to generate 19 additional mock data cubes, yielding the combined data volume that approximately cover the sky coverage of the HERA drift scan strip. Then, we repeat the simulation; that is, adding noise, wedge-cutting, averaging over frequency bins, and calculating the statistics. Results from all 20 fields are averaged together and the uncertainty propagated accordingly to obtain our final measurements.

\begin{figure*}
    \centering
    \includegraphics[width=0.95\textwidth]{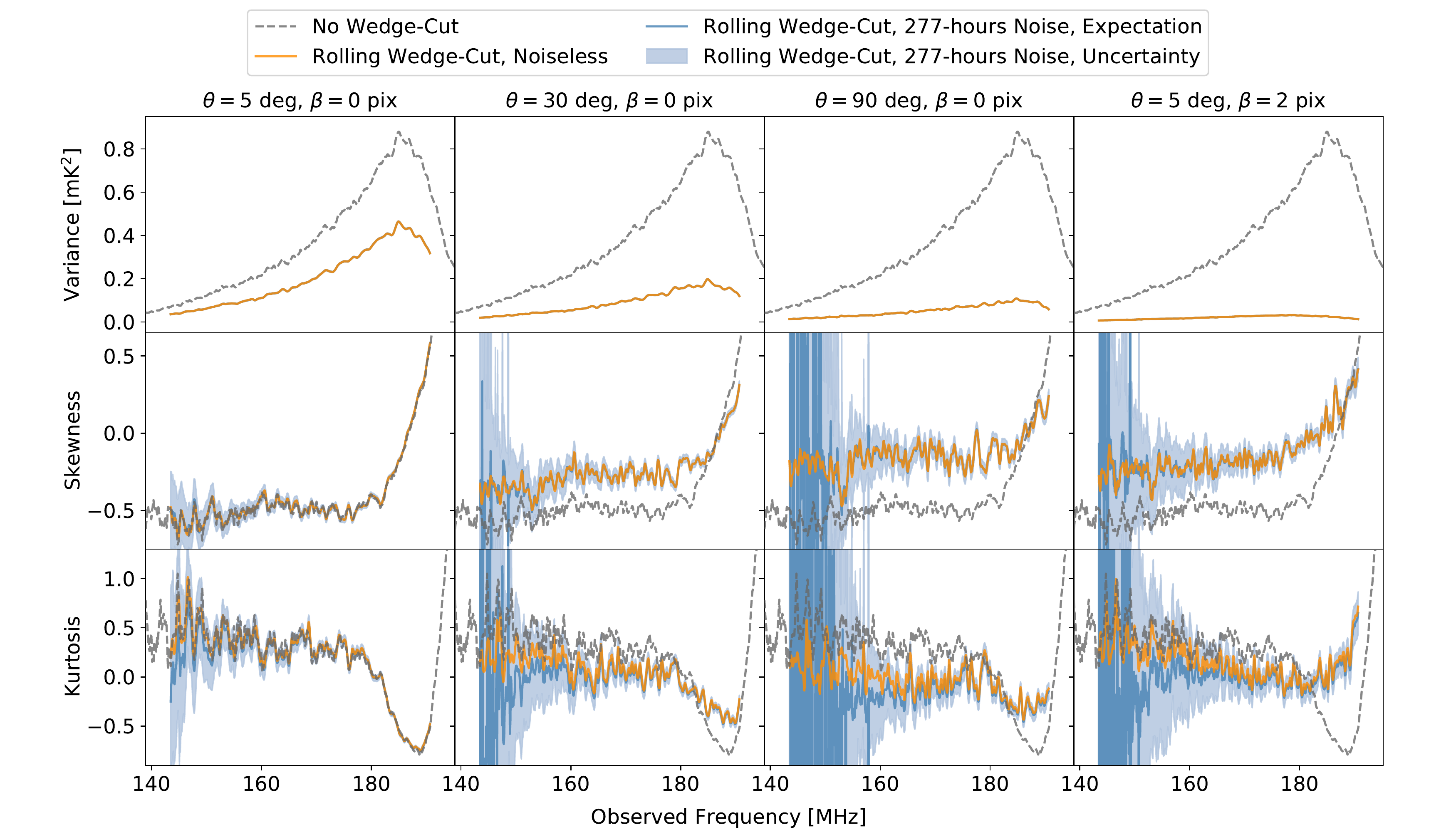}
    \caption{One-point statistics from the Monte Carlo simulations given the rolling wedge-cutting in Figure~\ref{fig:rolling_wc_maps_lightcone}. The calculations assume instrumental noise equivalent to 277 hours of integration time on the core of the full HERA array (331 antennas). Uncertainties includes both sample variances from 20 fields of observations and thermal uncertain derived from 1-$\sigma$ standard deviation from each Monte Carlo run. The intensity maps after rolling wedge-cut are averaged into 1-MHz bin to further improve the noise. As foreground wedge-cutting remove modes in the $k$ space that correspond to non-Gaussian structures, most of the non-Gaussian characteristics in the skewness and kurtosis of the 21~cm intensity maps disappear after wedge-cutting, yielding values that are more noise-like and fluctuate around zero. Nevertheless, the rises in skewness and kurtosis near the end of reionization appear to be less affected. The uncertainties of the variance are included in the plots, but they are very small and not visible.}
    \label{fig:rolling_wc_stats}
\end{figure*}

Figure~\ref{fig:rolling_wc_stats} shows the final statistics, assuming 277-hours of integration with 1~MHz frequency binning. As mentioned, the same rolling wedge-cut filters used to produce the intensity maps in Figure~\ref{fig:rolling_wc_maps_lightcone} are used to derive these statistics. The Monte Carlo simulations are done with 500 noise realisations. The expectation values from the simulations are shown in dark blue while the 1-$\sigma$ uncertainties are shown in light blue shade. Note that the uncertainties of the variance are very small and not visible in the plots. We also plot the measurements derived from the rolling wedge-cut noiseless intensity maps in orange for comparison. We focus on the specific noise figure (277 hours, 1-MHz bin) presented here because we find that it provides just enough sensitivity to make skewness and kurtosis detectable near the end of reionization. Without any frequency binning, the noise is extremely high, overwhelming the whole band, even with 1111 hours of integration. With a relatively small 1-MHz bin, coupled with 277 hours of integration, the lower 20-MHz of the band is still strongly dominated by noise, but the upper 10-MHz of the band gains significant sensitivity, enabling detection of the rises of skewness and kurtosis near the end of the EoR. Although we find in other test cases that binning over a larger bandwidth can further lower the noise level, redshift evolution appears to cause unexpected strong peaks and troughs that alter the overall trends of the statistics; thus, we choose to ignore those cases. Due to the strong noise level and loss of signal from wedge-cutting, we find it difficult to converge the Monte Carlo averages to the noiseless averages, especially in the lower 20-MHz of the band. We perform a test, where the number of noise realisations is slowly increased in the simulations, and find that 500 noise realisations are enough to obtain Monte Carlo derived skewness for the $\theta=90$ and $\beta=0$ case that is within 5\% of the noiseless average over the upper 10-MHz of the band. Thus, we terminate the Monte Carla simulations at 500 noise realisations for all cases, essentially choosing to prioritise test cases and paths of the frequency band that we think are important rather than spending significant more compute time.

The negative effects from wedge-cutting previously discussed in Section~\ref{sec:conventional_wc} still apply to measurements in Figure~\ref{fig:rolling_wc_stats} because the the rolling method does not reduces the bias from wedge-cutting on the statistics but only allows the full observing band (minus the edges) to be used. As the wedge-cutting becomes more aggressive, the variance reduces to near zero, the skewness turns into random fluctuation around zero, and the dip in kurtosis at $\sim185$ MHz also disappears rendering this feature undetectable. These directly follow from how wedge-cutting washes out ionised bubbles in the intensity maps in Figure~\ref{fig:rolling_wc_maps_lightcone}, which are the features that are responsible for the unique characteristic of the 21~cm one-point statistics. Without these bubbles, skewness and kurtosis become noisy. In regards to the different degrees of filtering, the first two cases in Figure~\ref{fig:rolling_wc_maps_lightcone} (counting from the left), though they look viable, correspond to minimal wedge-cutting that will not be enough to remove foreground contamination in real data. The third case that uses the wedge-cut filter with $\theta=90$ and $\beta=0$ corresponds to cutting the full foreground wedge without accounting for extra leakage. The rises in skewness and kurtosis at the end of the band still present with low uncertainty, making these features possible targets for detections. Once again, this case is still unrealistic because the use of a window function is unavoidable in the current data processing of 21~cm experiments. The realistic value of $\beta$ for the Blackman-Nuttall or Blackman-Harris window is $2-3$, but accounting for this will significantly remove statistical features as we have shown in the right most panel of Figure~\ref{fig:rolling_wc_maps_lightcone}. 
The last case in the fourth panel of Figure~\ref{fig:rolling_wc_stats}, which use a filter with $\theta=5$ and $\beta=2$, reiterates this point. That is, accounting for leakage from the Fourier transform window by shifting the wedge boundary to higher $k$ (setting $\beta>0$) is costly. This is apparent in the figure where the uncertainties on the statistics measured from the case with $\theta=5$ and $\beta=2$ are as high as (or higher than) the case with $\theta=90$ and $\beta=0$. This result reaffirms what we have seen in the conventional wedge-cut tests in Section~\ref{sec:wc_bias} that removing more large scale modes tend to remove more of the non-Gaussian features. However, as also discussed, it should not be interpreted as non-Gaussianity lives in low $k$ modes. The uncertainties on the statistics increase as the filtering becomes more aggressive, which suggests that the wedge-cut filter removes signal more than it removes thermal noise. Specifically, Gaussian noise power is flat across $k$ modes, and the noise being removed scales with the numbers of modes being removed when the wedge is cut, whereas much of the 21~cm signal lies in the low $k$ modes inside the wedge. Therefore, when more aggressive filters are used, most of the signals are removed but the noise level remains relatively unchanged, resulting in measurements that are more noise dominated. Nevertheless, our interest lies in the upper half of the band where skewness and kurtosis rise into strong signals, which, given the above mentioned noise level and binning, seem detectable in our most probable case ($\theta=90$ and $\beta=0$). In the future, we may be able to realistically perform this analysis with $\beta=0$ by implementing a different window function (see, e.g., \citealt{vedantham.etal.2012}). Alternatively, optimising the array layout can minimise the foreground wedge which would also reduce the power of the leakage from the window function (see, e.g., \citealt{murray.trott.2018}).

\begin{figure}
    \centering
    \includegraphics[width=\columnwidth]{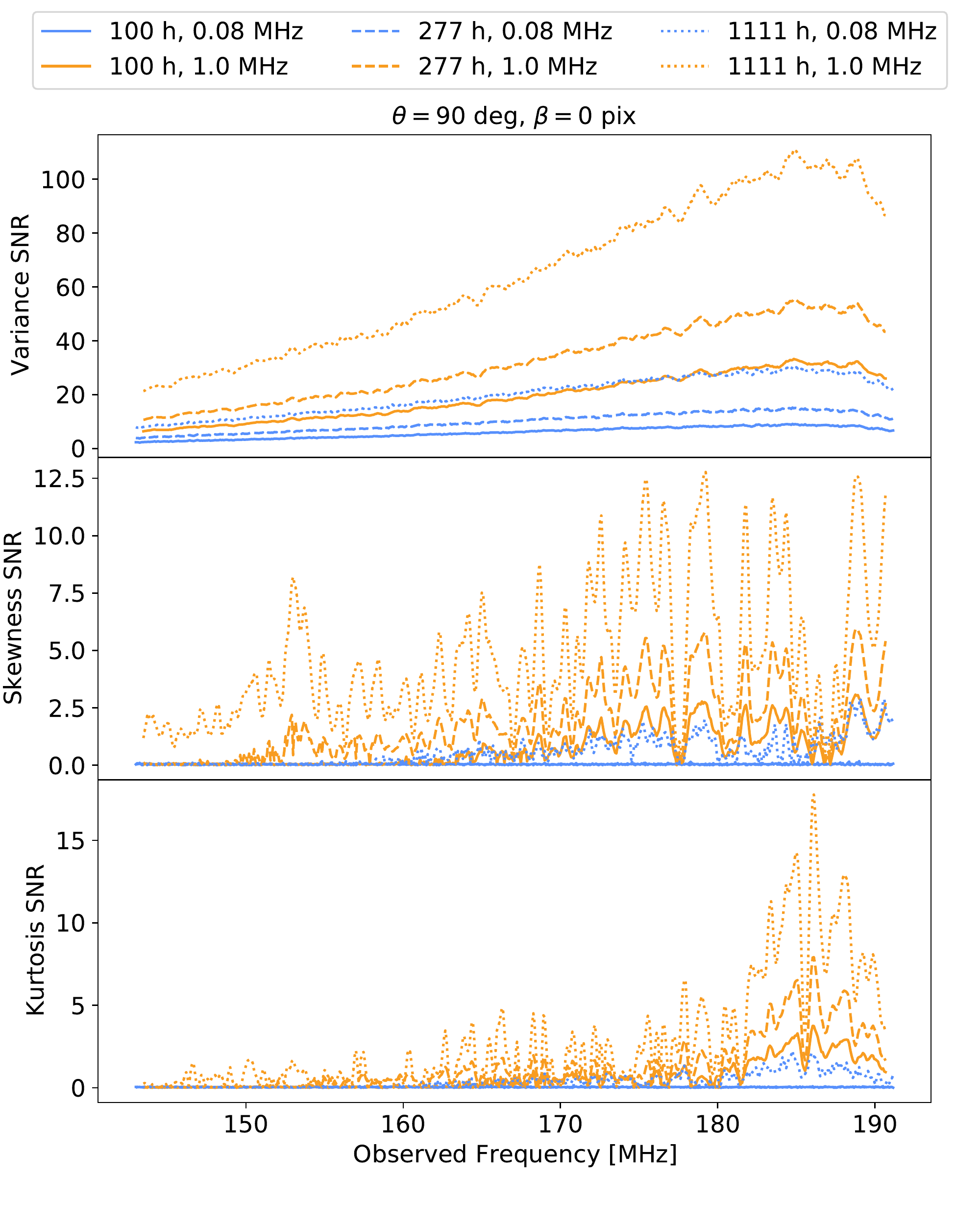}
    \caption{SNRs of the foreground wedge-cut one-point statistics derived from the Monte Carlo simulation given a full wedge-cut (assuming foreground sources from horizon-to-horizon) and the HERA instrument. We plot different combinations of integration time and frequency binning for comparison. These results demonstrate that measuring 21~cm one-point statistics through foreground avoidance is possible, but the numbers here do not mean to offer the absolute sensitivity number of the HERA instruments to these measurements.}
    \label{fig:rolling_wc_snr}
\end{figure}

To summarise, we calculate the SNRs, defined as the statistical values divided by 1-$\sigma$ uncertainties, and plot them in Figure~\ref{fig:rolling_wc_snr}. Once again, as our simulations do not include foregrounds and instrument systemics, these numbers only establish baseline expectations for the statistics with wedge removal strategy. We plot the SNRs from the no-binning (native 0.08-MHz bin) and 1-MHz binned cases given all three noise levels tested in the Monte Carlo simulations. The SNRs for the skewness and kurtosis are very noisy because the statistics themselves are noisy due to effects from foreground wedge-cutting. In some cases, e.g. at $\sim$178 MHz, the skewness and kurtosis values are near zero, yielding near zero SNRs. Nevertheless, we do not interpret these dips in the SNRs as random fluctuations because the overall trend still indicates increasing SNRs toward the end of reionization at the upper edge of the frequency band above $\goa$180 MHz where the rises in skewness and kurtosis still persist after the wedge removal. Among our test cases, we find 277 hours of integration together with 1-MHz frequency binning of the intensity maps after wedge-cutting to be a good compromising case that balance the SNR with integration time and noise reduction via frequency binning, but other combinations of integration time and frequency binning not shown in the figure can be used to achieve higher SNRs for detection as well. Note that it is not appropriate to sum over the SNRs in Figure~\ref{fig:rolling_wc_snr} over a range of frequency to get a total SNR across that range because the measurements at different frequencies are expected to be covariant due to the rolling filter.

\section{Conclusion}
\label{sec:conclusion}

In this paper, we have studied the effects of the foreground wedge-cutting on the 21~cm one-point statistics, following up on our previous work on the sensitivity analysis of HERA to these statistics in \citetalias{kittiwisit.etal.2018}. We constructed foreground wedge-cut filters based on our knowledge of the instrument mode-mixing foreground wedge and the spectral leakage from the Fourier transform window function. Then, we applied these filters to a mock data set to remove foreground contaminated modes and measure the statistics. We experimented with cutting the foreground wedge over different sizes of subbands and investigated various degree of wedge-cutting that included modifying the angle of the foreground wedge and shifting the wedge boundary to higher $k$.  

When performing wedge-cutting on a single subband, the window function weighting and un-weighting, which is essential to reduce ringing from Fourier transform operation, renders the edges of the subband unusable and makes it difficult to investigate the redshift evolution of the one-point statistics. Noticing that the middle of the subband is the least biased from wedge-cutting, we investigate a rolling wedge-cut method, where wedge-cutting is conducted on multiple subbands, and the map at the central frequency channel of the wedge-cut output from each subband is saved and stacked together to form the final foreground wedge-cut intensity cube. We apply an 8-MHz rolling wedge-cut with various degrees of wedge-cutting to our mock data cube and perform Monte Carlo simulations assuming three different levels of noise that correspond 100, 277, and 1111 hours of integration time on HERA. 

Overall, we found that wedge-cutting washed out the 21~cm intensity maps, removing the ionised regions. This has detrimental effects on the one-point statistics as these bubbles are responsible for the unique characteristics in the redshift evolution of the statistics. After wedge-cutting, the variance reduces to near zero, and the skewness and kurtosis become noisy and fluctuate around zero across the subband. The latter is discouraging as non-zero skewness and kurtosis can be used to quantify non-Gaussianity in 21~cm signals but rising trends in skewness and kurtosis at the end of reionization can be preserved in some cases. These effects are more severe when more $k$ modes are cut, especially the lower $k_{\parallel}$ modes. Our experiments with the various degrees of wedge-cutting shows that cutting more of the large scale modes tends to remove more of the non-Gaussian features, yielding 21~cm intensity maps that are more noise-like looking with washed out neutral islands and ionised bubbles, and skewness and kurtosis that fluctuate near zero. In addition, we find that more aggressive wedge-cut filters yield one-point statistics statistics with higher uncertainty in our Monte Carlo tests. This is expected as to the fact that the 21~cm power is concentrated in the lower $k_{\parallel}$ modes whereas the noise power is rather flat across the Fourier modes.

Nevertheless, our results suggest that measuring 21~cm one-point statistics via foreground wedge-cut strategy is possible if given sufficient integration time and some small frequency binning to improve SNR. Among our test cases, a full wedge-cut with no extra shift in the wedge boundary combined with the 277-hour noise and 1-MHz binning after foreground wedge-cutting preserves the rising of skewness and kurtosis signal near the end of reionization (panel 4 in Figure~\ref{fig:rolling_wc_maps_lightcone} and panel 3 in Figure~\ref{fig:rolling_wc_stats}) and yields the maximum SNRs $\approx 5$ for skewness and kurtosis. However, application of these results to upcoming HERA measurements is somewhat premature as our simulations do not include foregrounds or known instrumental systematics that may affect non-Gaussianity. Nevertheless, they shed light on paths towards measuring non-Gaussianity in HERA and similar experiments, as well as some of the challenges therein. 

The main finding from this work is that we must try to mitigate leakage from the Fourier transform window function through other means if we hope to recover one-point statistics (or other image-domain statistics) as cutting more of the higher $k$ modes to compensate for leakage outside of the wedge is very costly. One possible way to lower this leakage is to use a very large frequency subband, so that the window function spectrum is narrower in the $k_{\parallel}$, and thus less modes have to be cut. It may also be possible to reduce the foreground power in the wedge near the geometric horizon boundary by optimising the array layout \citep{murray.trott.2018} or the gridding of the visibility onto the $k$ space \citep{morales.etal.2019}. This would not remove the wedge but could lead to less foreground power being bled to the outside of the wedge by the window function, which could potentially allow cutting less modes above the geometric horizon of the wedge. Performing foreground subtraction, even partially, is another possible techniques to limit the leakage as lowering the wedge in general could help reduce its extent. Investigating these ideas will require a new set of simulations with good foreground models. An important improvement that can be done in our modelling (that is necessary to study the two optimisation ideas above) is to include the primary beam and PSF effect of the instrument. This will add chromatic behaviour of the beam, weighting the foreground sources differently across the sky, and produce the actual foreground wedge in the visibility. Another possibility for recovering not only one-point statistics but a wide range of image-based statistics from wedge-cut 21~cm images is machine learning. For example, a recent study by \citet{gagnon-hartman.etal.2021} has shown that machine learning algorithm can be used to recover some Fourier modes lost from wedge-cutting. A modification of such algorithm to target recovering of statistics could certainly be a promising approach for the future. We leave all these ideas for future studies. 

\section*{Acknowledgements}

PK thanks Phil Bull, and Mario Santos for supports and encouragements that are invaluable to the completion of this work, and acknowledges the financial assistance of the South African Radio Astronomy Observatory (SARAO; \url{www.sarao.ac.za}). We acknowledge support from the National Science Foundation through awards AST-1636646 and AST-1836019. BKG acknowledges the financial support from the National Science Foundation through an award for HERA (AST-1836019) and the European Research Council (ERC) under the European Union’s Horizon 2020 research and innovation programme (Grant agreement No.
884760, ``CoDEX'').

\section*{Data Availability}

The data underlying this article will be shared on reasonable request to the corresponding author.




\bibliographystyle{mnras}
\bibliography{references.bib} 


\bsp	
\label{lastpage}
\end{document}